\documentclass[
nofootinbib,
preprint,
showpacs,preprintnumbers,
amsmath,amssymb,
aps,
prd,
longbibliography,
floatfix,
lengthcheck,
]{revtex4-1}

\usepackage{graphicx}
\usepackage{dcolumn}
\usepackage{bm}
\usepackage{hyperref}
\usepackage{enumitem}
\usepackage{accents}

\DeclareMathOperator{\arccot}{arccot}
\DeclareMathOperator{\arctanh}{arctanh}
\DeclareMathOperator{\csch}{csch}

\begin{document}

\title{The initial data problem for a traversable wormhole with interacting mouths}

\author{Alexey L. Smirnov}%
\email{smirnov@inr.ac.ru}
\affiliation{Institute for Nuclear Research of the Russian Academy of Sciences,\\
60-th October Anniversary Prospect, 7a, 117312, Moscow, Russia
}

\date{\today}

\begin{abstract}
  In this study, we consider the time-symmetric initial data problem for GR minimally coupled with a phantom scalar field and a Maxwell field. The main focus is on initial data sets describing two interacting mouths of the {\it same} traversable wormhole. These data sets are similar in many respects to the Misner initial data with two black holes.
\end{abstract}

\pacs{04.20.Cv, 04.20.Fy}

\maketitle

\section{\label{sec:intro}Introduction and Preliminaries}
General Relativity and its possible modifications allow for the existence of Lorenzian traversable wormholes
if matter fields violate the null energy condition (NEC)~\cite{visser1,thorne1}.
The myriad of solutions obtained so far are mostly either
static spherically-symmetric or stationary axially-symmetric wormholes connecting
two {\it different} asymptotically flat universes. These are inter-universe dynamic wormholes according
to the terminology of~\cite{visser1}. 

On the other hand, one would try to construct traversable intra-universe wormholes.  
They connect regions of a single asymptotically flat universe and therefore their mouths can interact.
However, if such a wormhole spacetime is stable, then causality violations can become possible~\cite{thorne2}.   
In the present work wormholes are not expected to be stable.
Nevertheless, it is still of interest to study such spacetimes since the mouths of the wormhole
can be seen as an example of interacting exotic compact objects. Of course, finding the corresponding solutions
is now only possible numerically. Then the first step in this direction
is to solve the corresponding initial data problem (IDP), similarly to the case of binary black hole. 

In general, finding initial data in GR on a given~3-dimensional initial hypersurface~$\Sigma$
(usually called the Cauchy surface) requires solving four of the Einstein's equations, known as the constraint equations:
\begin{subequations}
\begin{eqnarray}
 ^{(3)}R+K-K_{ij}K^{ij} &=& 16\pi l^2_{\mathrm{P}}\rho\label{hc}\\
D_{j}(K^j_i-\delta^j_i K) &=& 8\pi l^2_{\mathrm{P}}S_i \label{mc},
\end{eqnarray}
\label{ceqs}
\end{subequations}
together with possible constraints imposed on matter fields.
In the equations $D_{\mu}$ and $^{(3)}R$ are respectively the covariant derivative and the scalar curvature
of the initial hypersurface, $K_{ij}$ is the extrinsic curvature of this hyperurface.
Matter sources of the gravitational field are characterized by a stress-energy tensor~$T_{\mu\nu}$ with energy density
~$\rho=T_{\mu\nu}n^{\mu}n^{\nu}$ and current density~$S_i=-T_{\lambda\mu}\gamma^{\lambda}_in^{\mu}$, where~$n^{\mu}$ is the unit timelike
normal of~$\Sigma$ and~$\gamma^{\lambda}_i$ is an operator of projection into~$\Sigma$.

The equations~\eqref{ceqs} can be simplified by the use of conformal techniques~\cite{york1,york2,york3}.
In this approach, the metric on~$\Sigma$ is assumed to be conformally equivalent to some prescribed ``background'' metric.
Then by carefully choosing conformal rescaling of the quantities in~\eqref{ceqs}, these equations can be decoupled and reduced to
a system of quasilinear elliptic equations.

The simplest illustration of the conformal approach arises when we consider the so-called time-symmetric initial data.
Time-symmetric initial data are obtained when there exists an initial slice with~\mbox{$K^i_j=0$}
and vanishing first time derivatives of the matter fields. Then the momentum constraint is identically zero
and one is left with the Hamiltonian constraint
\begin{equation}
\label{hcts}
^{(3)}R=16\pi l^2_{\mathrm{P}}\rho.
\end{equation}
plus possible constraints on matter fields. 

For example, if the metric on~$\Sigma$ is conformally flat then time-symmetric initial data sets for 
(electro)vacuum spacetimes can be given in terms of solutions of the flat space Laplace equation.
The corresponding harmonic functions were dubbed Lindquist as 'metric potentials'.
Such initial data sets describe multiple non-rotating back holes which are momentarily 
static~\cite{lindquist1,lindquist2,misner2}. 
 
In particular, the Brill-Lindquist initial data set 
with~$N$ black holes can be described by the initial slice~$\Sigma$ with~\mbox{$N+1$} asymptotically flat ends~\cite{lindquist2}
connected by~$N$ Einstein-Rosen bridges. 
Other examples of initial data sets were given by Misner~\cite{misner2} and Lindquist~\cite{lindquist1}. 
These data also describes~$N$ black holes
but this time~$\Sigma$ has only {\it two} asymptotically flat ends. By construction such Cauchy surface is invariant 
under inversion across throats of Einstein-Rosen bridges. Inversion-symmetric initial data for two black holes 
of equal mass also give rise to the so-called Misner wormhole~\cite{misner1} and its electrovac extension~\cite{lindquist1}. 
In these cases,~\mbox{$\Sigma$} can be made diffeomorphic to~\mbox{$\mathbb{S}^2\times\mathbb{S}^1$-\{point\}}.
Nevertheless, such initial data still describe a head-on collision of two black holes.

The paper considers the time-symmetric IDP for GR coupled with a phantom scalar field plus
possibly a Maxwell field. It appears that both Brill-Lindquist and Misner data can serve
as seed solutions for new metric potentials which solve the constraint equations.
The reversed sign of the kinetic term for the phantom field implies that new potentials must
be {\it complex} harmonic functions. In particular, complexification of the Misner
potential leads to initial data sets that solve the problem stated in the title.
Namely, if such data sets are regular, they describe two interacting mouths of a traversable wormhole momentarily at rest.

\section{\label{sec:spherical}The time-symmetric IDP for GR minimally coupled to a phantom scalar field}

\subsection{\label{sec:TS}General solution of the time-symmetric IDP.}
The theory of the real massless phantom scalar field minimally coupled with GR is defined by the equations
\begin{subequations}
\begin{eqnarray}
R_{\mu\nu} &=&-8\pi l^2_{\mathrm{P}}\nabla_{\mu}\phi \nabla_{\nu}\phi\label{eomm1},\\
\Box \phi &=& 0\label{eomf1}.
\end{eqnarray}
\label{eom1}
\end{subequations} 
As a starting point, consider Ellis-Bronnikov wormholes~\cite{ellis1,bronnikov1}
which are the following solutions of~\eqref{eom1}:
\begin{subequations}
\begin{eqnarray}
ds^2 &=& -e^{2by}dt^2 +e^{-2by}[dr^2 + (r^2+a^2)d\Omega^2],\quad\label{wm1}\\
\phi &=& \sqrt{\frac{1+b^2}{4\pi l^2_{\mathrm{P}}}}\,y,\quad y = \arccot \left(\frac{r}{a}\right),\label{wf1}
\end{eqnarray} 
\label{ellis1}
\end{subequations}
where~\mbox{$d\Omega^2=d\theta^2+\sin^2\theta d\varphi^2$} is the line element on 
the unit~2-sphere and~$b$, \mbox{$a>0$} are free parameters. 
The coordinate~\mbox{$r \in (-\infty, \infty)$} and spatial sections of the spacetime have 
two asymptotically flat sheets connected by the throat at~$r=-ab$. 
Since the theory~\eqref{eom1} is shift-symmetric, the scalar field is defined up to a constant.

The coordinate transformation 
\begin{equation}
r=\rho-a^2/4\rho\label{r2rho},
\end{equation}
with \mbox{$\rho \in (0,\infty)$} converts the solution~\eqref{ellis1} into the form
\begin{subequations}
\begin{eqnarray}
  ds^2 &=& -e^{2by}dt^2 +e^{-2by} \left( 1+\frac{a^2}{4\rho^2}\right)^2dl^2_{\flat}\label{wm2},\\
  dl^2_{\flat} &=& d\rho^2+\rho^2d\Omega^2,\\
  \phi&=&\sqrt{\frac{1+b^2}{4\pi l^2_{\mathrm{P}}}}\,y,\quad y=2 \arccot\left(\frac{2\rho}{a}\right).\\\label{wf2}
\end{eqnarray} 
\label{isot1}
\end{subequations}
The region~\mbox{$r>-ab$} is mapped into \mbox{$\rho_{\mathrm{th}}<\rho <\infty$}, the throat is at
\mbox{$\rho_{\mathrm{th}}=a(-b+\sqrt{1+b^2})/2$} and the region \mbox{$r<-ab$} is mapped into~\mbox{$0<\rho<\rho_{\mathrm{th}}$}. 
Thus any spatial sections of the spacetime~\eqref{wm2} in this coordinates is a punctured~$\mathbb{R}^3$.

Note when $b=0$,~\eqref{wm2} is isometric with respect to the inversion map
\begin{equation}
J:(\rho, \theta, \varphi) \mapsto (a^2/4\rho:=\tilde\rho,\tilde\theta=\theta,\tilde\varphi=\varphi).
\label{inv1}
\end{equation}
with the throat being its fixed point set. However, the scalar field is shifted by a constant under inversion.

The initial data problem~\eqref{hcts} for the theory~\eqref{eom1} is given by
\begin{equation}
  ^{(3)}R=-8\pi l^2_{\mathrm{P}}\partial_{i}\phi \partial^{i}\phi.
  \label{hctsph}
\end{equation}
Its general solution can now be obtained by means of inductive reasoning.
Using~\eqref{isot1} as a guiding example,~\eqref{isot1} leads to the crucial observation that the initial data for~\eqref{isot1}
can be defined in terms of the {\it unique} complex function~$\psi$.  Indeed, by making use of the complex
form of~$\arccot$, the initial data set for~\eqref{isot1} can be written as follows
\begin{subequations}
\begin{eqnarray}
  dl^2 &=& \psi^{2(1+ib)}\bar\psi^{2(1-ib)}dl^2_{\flat}=|\psi^{(1+ib)}|^4dl^2_{\flat},\label{sm3d}\\
\phi &=& i\sqrt{\frac{1+b^2}{4\pi l^2_{\mathrm{P}}}}\log \frac{\bar\psi}{\psi}+\mbox{const},\label{sf3d}
\end{eqnarray}
\label{sivd1}
\end{subequations}
The function~$\psi$ (``metric potential'') is the following complex-valued function 
\begin{equation}
\psi(\rho)=1+\frac{ia}{2\rho}\label{epsi}.
\end{equation} 

{\it However, the ansatz~\eqref{sivd1} will solve general time-symmetric IDP~\eqref{hctsph},
  provided~$\psi$ is any complex solution of the flat space Laplace equation
  \begin{equation}
    \Delta\psi=0
    \label{leq}
  \end{equation}
  such that the initial metric is asymptotically flat and positive and the initial field configuration is non-singular.}

\subsection{\label{sec:SS} Use the inversion map to glue data sets together: Spherically-symmetric example}
In spherical symmetry,~\eqref{leq} implies that the initial data set for the static wormhole is just one element
of a more general family of asymptotically flat initial data given by
\begin{equation}
\psi(\rho) =1 + \frac{A}{\rho},
\label{epsi1}
\end{equation}
where the parameter~\mbox{$A\in \mathbb{C}$}.
Note that metric potentials~\eqref{epsi1} no longer lead to static solutions for~\eqref{eom1}.
The remaining evolution equations must then be solved numerically. However, since the domain of integration
is a punctured~$\mathrm{R}^3$, one should somehow regularize the fields at the puncture by choosing appropriate boundary conditions.
However, among the functions~\eqref{epsi1} two specific potentials can be distinguished
\begin{equation}
  \psi_{\pm} = 1 + \frac{a e^{\pm\frac{i\lambda}{1+ib}}}{2\rho}\label{psiLR}.
\end{equation}
Here positive initial metric and non-singular~$\phi$ can be obtained in particular if~\mbox{$\lambda\in (-\pi,\pi)$}.

Individually the potentials~\eqref{psiLR} define two initial data sets
\begin{subequations}
\begin{eqnarray}
  dl_+^2 &=& |\psi_+^{(1+ib)}|^4dl^2_{\flat},\\
  \phi_+ &=& i\sqrt{\frac{1+b^2}{4\pi l^2_{\mathrm{P}}}}\log \frac{\bar\psi_+}{\psi_+}-\frac{\lambda}{ l_{\mathrm{P}}\sqrt{\pi(1+b^2)}},
\end{eqnarray}
\label{sivdL}
\end{subequations}
and
\begin{subequations}
\begin{eqnarray}
  dl_-^2 &=& |\psi_-^{(1+ib)}|^4dl^2_{\flat},\\
  \phi_- &=& i\sqrt{\frac{1+b^2}{4\pi l^2_{\mathrm{P}}}}\log \frac{\bar\psi_-}{\psi_-}.
\end{eqnarray}
\label{sivdR}
\end{subequations}
but  they are also transformed into each other by the inversion map~\eqref{inv1}.
This crucial property makes it possible to obtain a wormhole by properly gluing together these initial data sets along spheres with a radius of ~$a/2$.

For the vacuum case such gluing was described in the full generality by~\cite{misner2,lindquist1,giulini1}.
However, in order to explain how the simply connected geometries discussed in the next sections will turn into homotopically
non-trivial ones, it is appropriate to illustrate the procedure in the simplified setting of spherical symmetry.

Consider two manifolds (hereinafter called sheets),
both diffeomorphic to~$\mathbb{R}^3$ minus a ball of radius~$a/2$.
  Let the chart covering the first sheet be defined as
\begin{equation*}
\Sigma_- = \mathbb{R}^3-\mathbb{B}(a/2-\epsilon,0)
\end{equation*}
where~\mbox{$\mathbb{B}(a/2-\epsilon,0)$} is a ball of radius~\mbox{$a/2-\epsilon$}. The parameter~\mbox{$\epsilon>0$} defines
a outer collar neighbourhood of the boundary sphere. The coordinates are given by~\mbox{$\tilde\rho \in (a/2-\epsilon,\infty),\tilde\theta,\tilde\varphi$}. The initial data on the sheet are given by~\eqref{sivdR}.

The second sheet is covered by the chart
\begin{equation*}
\Sigma_+ = \mathbb{R}^3-\mathbb{B}(a/2,0)
\end{equation*}
with coordinates~\mbox{$\rho\in (a/2,\infty),\theta,\varphi$} and the initial data is ~\eqref{sivdL}.

The wormhole~$\Sigma$ is obtained by identifying  the open spherical shell
\begin{equation*}
\Delta_-(\epsilon) = \mathbb{ \mathring{B}}(a/2,0) - \mathbb{B}(a/2-\epsilon,0)
\end{equation*}
in the first chart, with the open spherical shell
\begin{equation*}
\Delta_+(\epsilon') = \mathbb{ \mathring{B}}(a/2+\epsilon',0) - \mathbb{B}(a/2,0)\nonumber
\end{equation*}
in the second chart by using the inversion map~\eqref{inv1}
\begin{equation*}
J|_{\Delta_+}: \Delta_+(\epsilon') \rightarrow \Delta_-(\epsilon)
\end{equation*}
Since~\eqref{inv1} is analytic away from the pole, the transition map~$J|_{\Delta_+}$ is analytic.

In the case of a time-symmetric slice, the throat of the wormhole is a minimal surface, see discussion in Appendix~\ref{sec:ap1}.
Thus for~$\Sigma$ it can be found by solving the equations
\begin{equation}
  D_is_{\pm}^i=0,
  \label{teq}
\end{equation}
in each~$\Sigma_{\pm}$ separately.  However, when~$b\neq 0$, one of these equations leads to a solution
which is always less than~$a/2$ and must therefore be excluded.
Thus there is only one minimal surface and the wormhole is asymmetric.

The evolution equations should also be solved separately on each sheet.
The evolving tensors on different sheets are related by
boundary conditions which are in fact similar to the so-called 'isometry boundary conditions' discussed in~\cite{alqubierre1}.
Such boundary conditions take into account the fact that~\eqref{inv1} is an involution,
therefore tensors may change their sign under the inversion.
Then the  boundary values of the tensors must be chosen in such a way as to avoid discontinuities at the boundary spheres.

\section{\label{sec:mult}Misner wormholes}
\subsection{\label{sec:vacuum}The Misner wormhole in vacuum}
It was mentioned in the introduction that the Misner initial data with two black holes of equal mass
can be treated as wormhole diffeomorphic to~\mbox{$\mathbb{S}^2\times\mathbb{S}^1$-\{point\}}. 
Since this solution plays a central role in the following, it deserves a separate discussion.

In the case of vanishing scalar field,~$\psi$ is real and the ansatz~\eqref{sivd1} is reduced to
\begin{equation}
  dl^2 =\psi^4dl^2_{\flat}=\psi^4(dx^2+dy^2+dz^2)\label{mm1}
\end{equation}

Then the first representation of the vacuum Misner data can be given as regular harmonic function in~$\mathbb{R}^3$ 
with two excised disjoint balls~$\mathrm{B}_{\pm}$ of same radius~\cite{misner2, lindquist1}. 
Let $a$ be the radius of the corresponding boundary 
spheres~$\mathrm{S}_{\pm}=\partial \mathrm{B}_{\pm}$ and $d$ is half the distance between their centres and the origin
of a cartesian coordinates system is at the midpoint between~$\mathrm{S}_{\pm}$.
We are looking for the solution of~\eqref{leq} with the following boundary conditions
\begin{subequations}
\begin{eqnarray}
  \frac{\partial\psi}{\partial n}+\frac{\psi}{2 a} &=& 0\quad \mbox{at}\quad S_{\pm}\label{bcond1},\\
  \lim_{|\mathbf{r}|\to\infty} \psi &=& 1\label{bcond2},
\end{eqnarray}
\label{mbvp1}
\end{subequations}
where~$\partial\psi/\partial n$ is the directional derivative along the outer normals to~$S_{\pm}$. Note, that the first
condition is just the minimal surface equation.

However, Misner proposed the method of constructing~$\psi$ which avoids solving the boundary value problem~\eqref{mbvp1} directly. The description of the method given below is a shortened version of the presentation given by Giulini~\cite{giulini1}.

First of all, we define the inversion maps with respect to~$S_{\pm}$.
\begin{equation}
\label{inv2}
J_{\pm}(\mathbf{r})=\pm\mathbf{d}+\frac{a^2}{|\mathbf{r}\mp\mathbf{d}|^2}(\mathbf{r}\mp\mathbf{d}).
\end{equation}
It is an involution~\mbox{$(J_{\pm})^2=id$} as expected. Now, there exists an involutive operator~${\cal J}_{\pm}$, which acts on 
functions defined on~$\mathbb{R}^3$ as follows
\begin{equation}
\label{J}
{\cal J}_{\pm}[f]=\frac{a}{|\mathbf{r}\mp\mathbf{d}|}f\circ J_{\pm}
\end{equation}
and having the property that for the Laplace operator:
\begin{equation}
\Delta \circ {\cal J}_{\pm} = \left(\frac{a}{|\mathbf{r}\mp\mathbf{d}|}\right)^4 {\cal J}_{\pm} \circ \Delta.
\end{equation}

Then the metric potential for the vacuum Misner wormhole is obtained by averaging the constant function~$f\equiv 1$ over the free
group generated by~$J_{\pm}$: 
\begin{equation}
\label{misnerpsi0}
\psi  = \left(1 + \sum_{n=1}^{\infty} \underbrace{{\cal J}_+{\cal J}_-{\cal J}_+\dots}_{\text{n factors}} + \sum_{n=1}^{\infty} \underbrace{{\cal J}_-{\cal J}_+{\cal J}_-\dots}_{\text{n factors}} \right)[\,1]
\end{equation}
With the use of~\eqref{J},~$\psi$ can be rewritten in the form
\begin{equation}
\psi=1+\sum_{n=1}^{\infty}a_n\left (\frac{1}{|\mathbf{r}+\mathbf{d}_n|}+\frac{1}{|\mathbf{r}-\mathbf{d}_n|}\right).
\label{misnerpsi1}
\end{equation}
Here~\mbox{$\pm{\mathbf d}_n=(0,0, \pm d_n)$} are the positions of the $n$-th poles inside~$\mathrm{S}_{\pm}$
and~$a_n$ is the $n$-th image charge
\begin{eqnarray}
d_1 &=& d\nonumber\\
d_n &=& d-\frac{a^2}{d+d_{n-1}}\qquad (n>1),
\label{d_n}
\end{eqnarray} 
while
\begin{eqnarray}
a_1 &=& a\nonumber\\
a_n &=& a_{n-1}\frac{a}{d+d_{n-1}}\qquad (n>1).
\label{a_n}
\end{eqnarray}
The series in~\eqref{misnerpsi1} are obviously positive and converge uniformly due to the Harnack's principle. 

Now in order to get a wormhole one should identify neighbourhoods of~$\mathrm{S}_{\pm}$. 
This procedure is similar to the one described in Section~\ref{sec:TS}. Let us introduce two spherical coordinate systems
centred at~$\pm{\mathbf d}_{1}$ and identify the corresponding collar neighbourhoods using the map~$J'$:
\begin{equation*}
  J':(r_-,\theta_-,\phi_-)\mapsto (r_+,\theta_+,\phi_+)=(a^2/r_-,\pi-\theta_-,\phi_-)
\end{equation*}
The subtle point here is that~$J'$ is now a composition of the inversion map with the orientation-reversing
diffeomorphism on~$\mathrm{S}_-$. This is necessary to get an orientable wormhole.

The second representation of the Misner wormhole is obtained by transforming~\eqref{misnerpsi1} and \eqref{mm1} to bispherical
coordinates~\mbox{$(-\infty<\mu<\infty,\,0\leq\eta\leq\pi,\,0<\phi\leq 2\pi)$}. 
Their relation to Cartesian coordinates is as follows
\begin{subequations}
\begin{eqnarray}
\label{bsc}
x &=& c\frac{\sin\eta}{\cosh\mu-\cos\eta}\cos\phi,\\
y &=& c\frac{\sin\eta}{\cosh\mu-\cos\eta}\sin\phi,\\
z &=& c\frac{\sinh\mu}{\cosh\mu-\cos\eta},
\end{eqnarray}
\label{bs}
\end{subequations}
Then the following line element is obtained from~\eqref{bs} and~\eqref{mm1}
\begin{equation}
dl^2=\hat\psi^4dl^2_D =\hat\psi^4(d\mu^2+d\eta^2+\sin^2\eta d\varphi^2)
\label{mm2}
\end{equation}
where 
\begin{equation}
\hat\psi=c^{\frac{1}{2}}\sum_{n=-\infty}^{\infty}\frac{1}{\sqrt{\cosh(\mu+2n\mu_0)-\cos\eta}}.
\label{misnerpsi2}
\end{equation}
The spheres~$\mathrm{S}_{\pm}$ are given by the equations~\mbox{$\mu=\pm\mu_0$} and therefore 
the coordinate~$\mu$ is restricted to the interval~\mbox{$[-\mu_0,\mu_0]$}.
The coefficient~\mbox{$c=\sqrt{d^2-a^2}$}.

The function~\eqref{misnerpsi2} is a solution of the elliptic equation
\begin{equation}
\Delta_D\hat\psi-\frac{1}{4}\hat\psi=0\label{elliptic} 
\end{equation}
with the Neumann boundary conditions
\begin{equation}
  \partial_{\mu} \hat\psi|_{\mu=\pm\mu_0}=0.
  \label{bcond3}
\end{equation}
The operator~$\Delta_D$ is the Laplase-Beltrami operator on~\mbox{$\mathbb{S}^2\times\mathbb{R}$}.

After identifying the spheres~$\mathrm{S}_{\pm}$, $dl^2_D$ becomes the metric on
a ``doughnut''~\mbox{$\mathbb{S}^2\times\mathbb{S}^1$}. 
Correspondingly,~\eqref{mm2} defines the metric on~\mbox{$\mathbb{S}^2\times\mathbb{S}^1$-\{point\}} 
since~$\hat\psi$ is {\it periodic} in~$\mu$ with period~$2\mu_0$.

\begin{figure}[tp]
\includegraphics[width=0.5\textwidth]{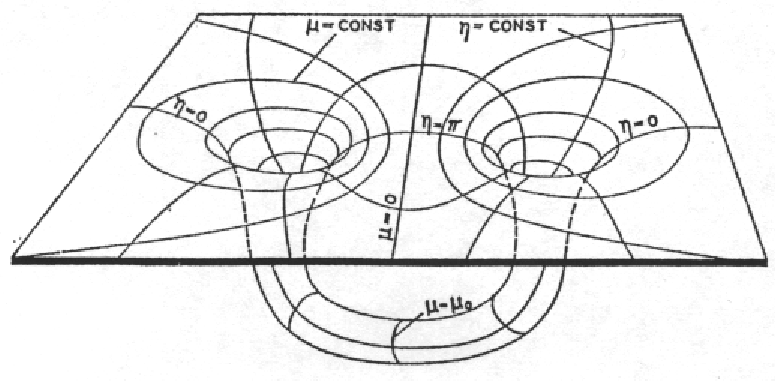}
\caption{\label{fig:misner} Embedding diagram for the Misner wormhole.}
\end{figure}

A visual picture of the wormhole can be obtained by the embedding in~$\mathbb R^3$ 
of two-dimensional section of the initial hypersurface with lines of constant~$\mu$ and~$\eta$
as shown in Fig.~\ref{fig:misner}. 

\subsection{\label{sec:ghost}Misner wormholes supported by the phantom scalar field}
If the scalar field is non-vanishing one can rewrite the ansatz~\eqref{sivd1} in bispherical coordinates
\begin{subequations}
\begin{eqnarray}
  dl^2 &=& |\hat\psi^{(1+ib)}|^4dl^2_D,\label{mmisnerm}\\
  \phi &=& i\sqrt{\frac{1+b^2}{4\pi l^2_{\mathrm{P}}}}\log \frac{\bar{\hat\psi}}{\psi}+\mbox{const},\label{mmisnerf}
\end{eqnarray}
\label{mmids}
\end{subequations}
and require two conditions to be met:
\begin{enumerate}[label=(\roman*)]
\item Similarly to the vacuum case, the metric must be periodic
  \begin{equation}
    |\hat\psi(\mu+2\mu_0,\eta)^{(1+ib)}|^4=|\hat\psi(\mu,\eta)^{(1+ib)}|^4.\label{periodic}
  \end{equation}
\item At the same time, if~$\mu$ is shifted by~$2\mu_0$, the scalar field must be transformed as follows
  \begin{equation}
    \phi(\mu+2\mu_0,\eta)=\phi(\mu,\eta)-\frac{\lambda}{l_{\mathrm{P}}\sqrt{\pi(1+b^2)}}.\label{multiv}
  \end{equation}
\end{enumerate}
The latter condition says that~$\phi$ is a multivalued function. However, since the zero mode of the free scalar field
is physically insignificant, multivaluedness is harmless. Note also that~\eqref{periodic} and~\eqref{multiv}
are direct counterparts of the "match-up" conditions introduced by Lindquist~\cite{lindquist1} for electrovacuum wormholes.  

Now~\eqref{periodic} and~\eqref{multiv} along with~\eqref{misnerpsi2} uniquely determine the function~$\hat\psi$ as follows
\begin{equation} 
\hat\psi = c^{\frac{1}{2}}\sum_{n=-\infty}^{\infty}\frac{e^{\frac{in\lambda}{1+ib}}}{\sqrt{\cosh(\mu+2n\mu_0)-\cos\eta}}
\label{mmisnerpsi1}
\end{equation}
and the corresponding expression in cartesian coordinates is given by
\begin{equation}
 \psi = 1+\sum_{n=1}^{\infty}a_n\left(\frac{e^{\frac{in\lambda}{1+ib}}}{|\mathbf
{r}+\mathbf{d}_n|}+\frac{e^{-\frac{in\lambda}{1+ib}}}{|\mathbf{r}-\mathbf{d}_n|}
\right)\label{mmisnerpsi2} 
\end{equation}
This is the main result of this section. These solutions will still be referred to as Misner wormholes thereafter.  

The ratio test shows that the series~\eqref{mmisnerpsi2} converge absolutely when
\begin{equation}
\frac{|\lambda b|}{1+b^2} < \mu_0\label{conv}\quad\mbox{with}\quad\lambda\in(\pi,\pi).
\end{equation}
and hence it is also convergent under these conditions. Another way to get~\eqref{conv} is to 
decompose~$\hat\psi$ into the sum of two general Dirichlet series and calculate their abscissae of
absolute convergence. The convergence of~\eqref{mmisnerpsi2} implies of course that~\eqref{mmisnerpsi2}
is also convergent. 

In order to have strictly positive metrics and non-singular~$\phi$, the function~$\hat\psi$ must have
no zeros outside of~$\mathrm{S}_{\pm}$. Note that  potentials with~\mbox{$b=0$} and ~\mbox{$\lambda=\pm\pi$}
violate this condition at~\mbox{$\mu=\pm\mu_0$} and were therefore excluded.
Numerical investigations show that there are probably no zeros for~$\hat\psi$ outside of~$\mathrm{S}_{\pm}$ for other
values of the parameters, but the exact proof is currently lacking.

In the limiting case where the mouths of the wormhole are well separated i.e. when~\mbox{$d\gg a$}, one can expand
the potential~\eqref{mmisnerpsi1} in powers of~$a/d$ using  cartesian coordinates centred at one of the mouths.
Then taking into account~\eqref{d_n} and~\eqref{a_n} one obtains
\begin{equation}
\psi=1 + \frac{a e^{\pm\frac{i\lambda}{1+ib}}}{|\mathbf{r'}|} + O\left(\frac{a}{d}\right)
\end{equation}
where~\mbox{$\mathbf{r'} = \mathbf{r} \pm \mathbf{d}$}.
Thus the leading order of the expansion coincides with~\eqref{psiLR}. Therefore, each individual mouth can be described
as $\pm$ initial data sets discussed in Section~\ref{sec:SS}.  In particular, initial states of
Ellis-Bronnikov wormholes can now be treated as the long-range limit of Misner wormholes when~$b=0$. 

The throat in~\eqref{mmisnerpsi1} can only be seen explicitly for the case~\mbox{$b=0$}.
It is still  the surface~\mbox{$\mu=\pm\mu_0$}. Indeed, the minimal surface equation is reduced to
\begin{equation}
\partial_{\mu}(\hat\psi\bar{\hat\psi})|_{\mu=\pm\mu_0}=0.
\label{mmisnerteq1}
\end{equation}
which is satisfied since the following relations hold
\begin{subequations}
\begin{eqnarray}
  \bar{\hat\psi}|_{\mu=\pm\mu_0} &=& e^{\pm i\lambda}\hat\psi|_{\mu=\pm\mu_0},\\
  \partial_{\mu}\bar{\hat\psi}|_{\mu=\pm\mu_0} &=&-e^{\pm i\lambda}\partial_{\mu}\hat\psi|_{\mu=\pm\mu_0}.
\end{eqnarray}
\end{subequations}

A few comments are in order regarding~\eqref{mmisnerteq1}. We could treat this expression
in a similar way to~\eqref{bcond3}, i.e. as a boundary condition. Note that while~\eqref{bcond3}
is a well-known Neumann boundary condition,~\eqref{mmisnerteq1} is a {\it nonlinear} boundary condition.
However, since~\eqref{elliptic} is linear and solutions can be constructed
using only the "match-up" conditions~\eqref{periodic} and~\eqref{multiv}. 
On the other hand, the ansatz~\eqref{mmids} could be applied to some other slicing.
Then the constraint equations would lead to some quasilinear elliptic equation for complex~$\psi$.
If the slices contain minimal surfaces, then one would inevitably have to deal with 
the nonlinear boundary value problem, where the boundary conditions are given by~\eqref{mmisnerteq1} 
because it is not clear how to implement~\eqref{periodic} and~\eqref{multiv} in the case of
non-linear equations.

When~\mbox{$b\neq0$}, the surface~$\mu=\pm\mu_0$ is no longer minimal. Rather, the minimal surface is now a surface of
revolution with generatrix~\mbox{$\mu(\eta)$}. It can only be found numerically.

Further, it follows from  the Raychaudhuri equation that minimal surfaces are indeed throats of initially traversable wormholes.
In fact, this equation defines the "flare-out" conditions on the throat.
The detailed discussion is given in Appendix~\ref{sec:ap1}.

\begin{figure}[tp]
\includegraphics[width=0.4\textwidth]{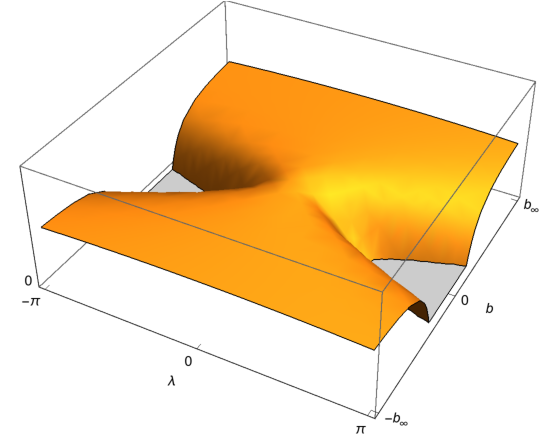}
\caption{\label{fig:mADM} The positive part of~$m_{\mathrm{ADM}}(\lambda,b,\mu_0)$ for the region
  where~\mbox{$\lambda\in (-\pi,\pi)$},~\mbox{$b\in [-b_{\infty},b_{\infty}]$} and some~\mbox{$\mu_0$=const}.
  When~\mbox{$|b_{\infty}|\to\infty$}, the function approaches to a constant value.}
\end{figure}

It is known that perturbed static wormholes supported by a phantom scalar field
are unstable~\cite{hayward2,gonzalez2,gonzalez3}. The instability turns them into so-called dynamic wormholes~\cite{hayward1}.
Since Ellis-Bronnikov wormholes emerge from initial data sets~\eqref{mmisnerpsi1} in the limit of large distances,
there is reason to suspect that spacetimes developing from regular~\eqref{mmisnerpsi1} are also unstable.
However, if the ADM masses of such spacetimes are positive, then the end state of the evolution is a black hole,
regardless of stability issues. Note, that the expression for the ADM mass is given by 
\begin{widetext}
\begin{equation}
  m_{\mathrm{ADM}}= \frac{c}{l^2_{\mathrm{P}}}\sum_{n=1}^{\infty}\frac{1}{\sinh n \mu_0}\left\{ \cos\left(\frac{\lambda n}{1+b^2}\right)\cosh\left(\frac{\lambda b\,n}{1+b^2}\right)-b \sin\left(\frac{\lambda n}{1+b^2}\right)\sinh\left(\frac{\lambda b\,n}{1+b^2}\right)\right\}
\end{equation}
\end{widetext}
We can then verify that wormholes of positive mass do indeed exist in the theory by looking at the positive part
of the function~$m_{\mathrm{ADM}}(\lambda,b,\mu_0)$. Its typical form is shown in Fig.~\ref{fig:mADM}.

To conclude this section, note that the time scales associated with instabilities are expected to be very small,
as the case of spherically-symmetric wormholes shows.
However, the addition of an electromagnetic charge may potentially serve as a stabilization mechanism.
The aim of the next section is to extend the theory to include an electromagnetic field
and study wormhole solutions of the corresponding IDP.

\subsection{\label{sec:charged}Addition of a Maxwell field to the theory.}
In this case, the equation of motions are given by   
\begin{subequations}
\begin{eqnarray}
  R_{\mu\nu} &=&-8\pi l^2_{\mathrm{P}}\nabla_{\mu}\phi \nabla_{\nu}\phi\\
    & &+2 l^2_{\mathrm{P}}\left(F^{\lambda}_{\mu}F_{\nu\lambda}-\frac{1}{4}g_{\mu\nu} F_{\alpha\beta}F^{\alpha\beta}\right) \label{eomm3},\\
\Box \phi &=& 0\label{eomf3},\\
\nabla_{\mu}F^{\mu\nu}&=& 0\label{eommf3}.
\end{eqnarray}
\label{eom3}
\end{subequations}
For simplicity only the case of electric field will be considered. 
The general time-symmetric IDP for~\eqref{eom3} is the system of hamiltonian and Gauss constraints
\begin{subequations}
\begin{eqnarray}
  ^{(3)}R &=&-8\pi l^2_{\mathrm{P}}\partial_{i}\phi \partial^{i}\phi+2 l^2_{\mathrm{P}} E_iE^i,\label{hctsphE}\\
  D_iE^i &=& 0\label{gauss},
\end{eqnarray}
 \label{charged_idp}
\end{subequations}
where~$E_i$ is the electric field in the initial hypersurface.

As expected, the simplest wormhole initial data can be learned from static spherically-symmetric wormhole solutions.
For the purposes of the present paper, it is interesting to consider the following  Matos-Gonz\'{a}lez-Guzm\'{a}n-Sarbach
(MGGS) solution~\cite{matos1,gonzalez1}
\begin{subequations}
\begin{eqnarray}
  ds^2 &=& -U^{-2}dt^2 +U^2\left( 1+\frac{a^2}{4\rho^2}\right)^2dl^2_{\flat}\label{wm3},\\
  \phi &=&\sqrt{\frac{1+C^2}{4\pi l^2_{\mathrm{P}}}-\frac{{\cal Q}^2}{4\pi a^2}}\,y,\label{wf3}\\
  F&=&\frac{{\cal Q}}{(U\rho)^2\left( 1+\frac{a^2}{4\rho^2}\right)}\,dt\wedge d\rho,\label{wef3}\\
  y &=& 2 \arccot\left(\frac{2\rho}{a}\right),\nonumber\\
   U &=& \cos\left(\Omega\,y\right)-C\,\frac{\sin\left(\Omega\,y\right)}{\Omega},\,\, \Omega=\sqrt{\frac{ l^2_{\mathrm{P}}{\cal Q}^2}{a^2}-C^2}.\nonumber
\end{eqnarray} 
\label{isot2}
\end{subequations}
This solution belongs to the supercritical class (in the terminology of~\cite{gonzalez1}) since its ADM mass is positive.
In addition, to have a globally well-defined wormhole, the parameters must be constrained by conditions
\begin{equation}
  \frac{C}{\Omega}\tan(\Omega\pi)<1\quad \mbox{and}\quad \Omega<\frac{1}{2}\label{pcond1}.
\end{equation}
The time-symmetric initial data set for~\eqref{isot2} will serve as useful reference example in the further discussion. 

Interestingly, it is possible to find a solution
of~\eqref{charged_idp} if a seed solution of~\eqref{hctsph} is already known. Originally a similar observation was made
by Ort\'{i}n~\cite{ortin1} to solve the constraint equations for dilaton gravity.

Let a solution of~\eqref{hctsph} be given by an initial configuration of~$\phi$ and an initial metric such that
\begin{subequations}
\begin{eqnarray}
  dl^2 &=& W^2dl^2_{\flat},\label{seedm}\\
  D_iD^i\phi &=& 0,\label{phieq}
\end{eqnarray}
\label{uncharged_ids}
\end{subequations}
The latter condition severely restrict the available seed solutions.  
In particular, it only hold for the initial data sets considered earlier in the paper if the parameter~\mbox{$b=0$}.

Accordingly, let a solution of~\eqref{charged_idp} be provided by
\begin{subequations}
\begin{eqnarray}
  d\tilde l^2 &=& \tilde W^2dl^2_{\flat},\\
  \tilde\phi &=& B\phi\\
  E_i &=& -\partial_iZ,\\
  Z &=& f(\phi).
\end{eqnarray}
\label{charged_ids}
\end{subequations}
Here~$B$ is a constant and~$E_i$ has been expressed through the electric potential~$Z$.

Now,~\eqref{charged_idp} can be reduced to a single equation for the unknown function~$f(\phi)$.
First of all, from the Gauss constraint and~\eqref{phieq} we get the relation
\begin{equation}
  \tilde W = W\,(f^{\prime})^{-1},\quad f^{\prime}:=\frac{df}{d\phi}\label{tildeW}.
\end{equation}
Then substituting~\eqref{charged_ids} and~\eqref{tildeW} into~\eqref{hctsphE} and taking into account~\eqref{phieq} we get
\begin{equation}
  3(f^{\prime\prime})^2+2 f^{\prime}f^{\prime\prime\prime}+p^2(f^{\prime})^2+q^2(f^{\prime})^4=0\label{feq}\end{equation}
where~\mbox{$p^2=4\pi l^2_{\mathrm{P}}(1-B^2)$} and~\mbox{$q^2=2 l^2_{\mathrm{P}}$}. The particular solution of this equations is
\begin{equation}
  f(\phi)=\frac{2}{q}\arctanh\left[\frac{C_1+p\,\tan\left(\frac{p\,\phi}{2}+C_2\right)}{q}\right],\label{fE1}
\end{equation}
where~\mbox{$C_1^2=q^2-p^2$}.
Then using the Ellis-Bronnikov ($b=0$) and MGGS initial data sets as~\eqref{uncharged_ids} and~\eqref{charged_ids}
respectively, one can fix the remaining constants~$B$ and~$C_2$.

Now one could try to construct a Misner initial data using the function~\eqref{fE1} and the seed solution
defined by the metric potential~\eqref{mmisnerpsi1}.
Then the corresponding solution of~\eqref{charged_idp} in bispherical coordinates
\begin{eqnarray}
  dl^2 &=& \left[\cos(p\,\phi)-\frac{C}{\Omega}\sin(p\,\phi)\right]^2|\hat\psi|^4dl^2_D,\label{mmisnerE}\\
  \tilde\phi &=& B\phi,\quad\phi=\frac{i}{\sqrt{4\pi l^2_{\mathrm{P}}}}\log\frac{\hat\psi}{\bar{\hat\psi}},\label{fminsnerE}\\
  E_i &=&-Q\left[\cos(p\,\phi)-\frac{C}{\Omega}\sin(p\,\phi)\right]^{-1}\partial_i\phi,\label{emisnerE}
\end{eqnarray}
This solution would give a well-defined wormhole initial data if (i) the expression in square the brackets is positive
and (ii) the metric is periodic with period~$2\mu_0$. Obviously this is impossible. Therefore, unfortunately
the MGSS solution cannot be considered as a large distance limit for a Misner wormhole.

Nevertheless, if~\mbox{$B^2>1$} there exists a particular solution
\begin{equation}
  f(\phi)=\sqrt{2\pi (B^2-1)}\,\phi,
\end{equation}
where the constant~$B$ can be chosen in such a way that~\mbox{$\sqrt{2\pi (B^2-1)}=1$}.
Thus we conclude that symmetric Misner wormholes with~\mbox{$b=0$} are also solutions of~\eqref{charged_idp}.
Now, by applying the transformation~\mbox{$\lambda\to i\lambda$},~\eqref{mmisnerpsi1}
can be formally obtained from metric potentials given in~\cite{lindquist1}.
This implies that the charges associated with different sides of the throat can be written as
\begin{equation}
  q_1=-q_2=\frac{2 c}{l_{\mathrm{P}}}\sum_{n=1}^{\infty}\sin\,n\lambda \csch\,n\mu_0
\end{equation}
Similarly, formal expressions could be given for the masses~$m_1$ and~$m_2$ of each mouth.
However, the definition of quasi-local mass used by Lindquist was questioned in~\cite{giulini1}.
Therefore the meaning of such expressions for~$m_1$ and~$m_2$ remains unclear. Instead, the Penrose's quasi-local mass
should probably be used. Its calculation is a non-trivial task even for the vacuum wormhole~\cite{tod1}
and is left for future work.

In conclusion, it should be noted that the method proposed for solving the IDP~\eqref{charged_idp}
can also be applied to non-linear electrodynamics. 

\section{\label{sec:disc}Conclusions and outlook}
In the present paper several extensions of the Misner initial data have been obtained.
The corresponding initial data sets are solutions of the constraint equations for GR minimally coupled 
to a free massless phantom scalar field and a Maxwell field at the moment of time symmetry.
These solutions are defined by a single complex harmonic function, which appears to be the Misner
harmonic function with complex image charges. In retrospect, solutions presented in the paper can also be seen as a
complexification of the corresponding initial data sets obtained in~\cite{ortin1} for the canonical scalar field.

The constructed data sets were interpreted as pairs of interacting mouths of a dynamic wormhole.
If the ADM mass of such a spacetime is positive, one can expect that scalar hairs will be radiated away and/or
collapse during time evolution and an initially traversable wormhole will inevitably be sealed.
Eventually the mouths collapse into a final black hole.
However, unlike the case of head-on collision of two vacuum black holes,
the actual picture is complicated by the fact that the phantom scalar field exerts a repulsive force. In addition,
fluctuations of the field are likely to make the wormhole unstable. But of course, the ultimate judge of possible scenarios
would be numerical relativity simulations.

While the paper described basic features of initial data sets, it left many questions unanswered.
In particular, the formal proof of the regularity of data sets has not been given. Also, while the ADM mass can be easily
obtained from data sets, the individual masses of mouths are more difficult to obtain. It seems that the correct way
of doing this is to calculate the Penrose's quasi-local mass. 

The results of the paper can be further extended. As an obvious suggestion, one can try to use ansatz~\eqref{sivd1}
with different slicing conditions. For example, in the case of the maximal slicing~\mbox{$K=0$},
by using~\eqref{sivd1} with~\mbox{$b=0$}, it is possible to reduce the constraint equations to a single
quasilinear equation for the complex function~$\psi$. However, if slices contains minimal surfaces (throats), the corresponding
boundary value problem becomes non-linear.

All constructed initial data sets can be also used when positive cosmological constant~$\Lambda$ is added.  
It was shown in~\cite{maeda1} that solutions of the time-symmetric IDP without the cosmological constant
are also solutions of the IDP with non-vanishing~\mbox{$\Lambda>0$} provided  that the slicing condition is given
by~\mbox{$K_i^j=-\sqrt{\Lambda/3}\,\delta_i^j$}.
In fact such slicing arises naturally in GR coupled with a complex scalar field and spontaneously broken global~$U(1)$ symmetry.
While the topology of the initial surface is still~\mbox{$\mathbb{S}^2\times\mathbb{S}^1$-\{point\}},
the minimal surface is no longer a trapping surface (this can be seen from~\eqref{lescalar} and~\eqref{kescalar})
and the notion of wormhole throat becomes ambiguous.

\begin{acknowledgments} 
The author is grateful to Sergey Mironov, Dmirtry Levkov, Sergey Demidov and Victor Berezin for useful discussions.
\end{acknowledgments}
 
\appendix
\section{\label{sec:ap1}Flare-out conditions on the throat}
The fact that the throat is given by the minimal surface in a time-symmetric Cauchy slice
was used in the main text to establish the presence of wormholes in initial data sets. 
However, traversability of the wormhole cannot be deduced solely form the intrinsic geometry of the initial slice.
We must study the spacetime in the vicinity of the minimal surface.

Consider then a closed orientable two-dimensional surface~$\cal{S}$ embedded in some
(at the moment not necessarily time-symmetric) Cauchy slice~$\Sigma$. Let~$s_{\mu}$ be the spacelike outward-pointing
normal to~$\cal{S}$ and~$n_{\mu}$ the timelike unit future-pointing normal to~$\Sigma$. By construction,
these vectors are orthogonal to each other:~\mbox{$s_{\mu}n^{\mu}=0$}.
Using these vectors, two additional future-pointing null vector fields can be defined
\begin{eqnarray}
  l_{\mu}&=&\frac{1}{\sqrt{2}}(n_{\mu}+s_{\mu})\\
  k_{\mu}&=&\frac{1}{\sqrt{2}}(n_{\mu}-s_{\mu}).
\end{eqnarray}
By construction, they are normed as follows:
\begin{equation}
l_{\mu}l^{\mu}=k_{\mu}k^{\mu}=0,\quad l_{\mu}k^{\mu}=-1
\end{equation}
and represent "outgoing" $l$ and "ingoing" $k$ null vector fields orthogonal to~$\cal{S}$.
Therefore $l$ and $k$ can also define tangent vectors to null geodesics composing respectively outgoing
and ingoing null congruences near~$\cal{S}$.

Then the metric on~$\cal{S}$ is given by 
\begin{equation}
  q_{\mu\nu}=g_{\mu\nu}+l_{\mu}k_{\nu}+k_{\mu}l_{\nu}=\gamma_{\mu\nu}-s_{\mu}s_{\nu}
  \label{qmetric}
\end{equation}
In the first equality~$ q_{\mu\nu}$ is considered as the metric induced by the spacetime
metric~$g_{\mu\nu}$ while in the second equality the same metric induced by the spatial
metric~$\gamma_{\mu\nu}$ in~$\Sigma$. 

Now, consider the pair of expansion tensors for each null direction
\begin{widetext}
\begin{eqnarray}
  \theta^{(l)}_{\mu\nu}&=&q^{\alpha}_{\mu}q^{\beta}_{\nu}\nabla_{\alpha}l_{\beta}=\frac{1}{\sqrt{2}}(D_{\mu}s_{\nu}-K_{\mu\nu}+s_{\mu}s^{\lambda}K_{\lambda\nu}-s_{\mu}s^{\lambda}D_{\lambda}s_{\nu})\label{letensor}\\
  \theta^{(k)}_{\mu\nu}&=&q^{\alpha}_{\mu}q^{\beta}_{\nu}\nabla_{\alpha}k_{\beta}=\frac{1}{\sqrt{2}}(-D_{\mu}s_{\nu}-K_{\mu\nu}+s_{\mu}s^{\lambda}K_{\lambda\nu}+s_{\mu}s^{\lambda}D_{\lambda}s_{\nu})\label{ketensor}
\end{eqnarray}
\end{widetext}
The form of expansion tensors in terms of geometric data given in~$\Sigma$ has been obtained by using~\eqref{qmetric}
and the following relations
\[
K_{\mu\nu}=-\gamma^{\alpha}_{\mu}\gamma^{\beta}_{\nu}\nabla_{\alpha}n_{\beta},\quad D_{\mu}s_{\nu}=\gamma^{\alpha}_{\mu}\gamma^{\beta}_{\nu}\nabla_{\alpha}s_{\beta}
\]
Now, the expansion scalars (or simply expansions) are obtained by contracting~\eqref{letensor} and~\eqref{ketensor}
with the metric~\eqref{qmetric}:
\begin{eqnarray}
  \theta^{(l)}&=& \frac{1}{\sqrt{2}}(D_{\mu}s^{\mu}+K_{\mu\nu}s^{\mu}s^{\nu}-K)\label{lescalar}\\
  \theta^{(k)}&=& \frac{1}{\sqrt{2}}(-D_{\mu}s^{\mu}+K_{\mu\nu}s^{\mu}s^{\nu}-K)\label{kescalar}
\end{eqnarray}
The expansion scalars can be treated as the fractional rate of change of the cross-sectional
area of the corrsponding null geodesic congruence. Heuristically the expansion is negative if the light rays are converging,
otherwise it is positive.

The surface~$\cal{S}$ is called untrapped if~$\theta^{(l)}$ and~$\theta^{(k)}$ have opposite signs.
If both expansions are negative or positive then~$\cal{S}$ is called trapped or anti-trapped respectively.
The surface~$\cal{S}$ is marginal when either one or both of the expansions are zero.
A hypersurface foliated by marginal surfaces is called a trapping horizon.
Further classification of boundary surfaces is discussed in detail in~\cite{yang1}. 

Now, if~$\Sigma$ is time-symmetric,~\mbox{$K_{\mu\nu}=0$} and it is follows
from~\eqref{lescalar} and~\eqref{kescalar} that the throat~$\cal{S}$ of an initial wormhole
is simultaneously a degenerate marginal surface and a minimal surface in~$\Sigma$ i.e.
\[
\theta^{(l)}=\theta^{(k)}=0\quad \text{iff}\quad D_is^i=0 
\]
as one would expect.

By definition, the hypersurfaces~\mbox{$\theta^{(l)}=0$} and~\mbox{$\theta^{(k)}=0$} are trapping horizons.
Then the normals to these hypersurfaces are given by~$\nabla_{\mu}\theta^{(l)}$ and~$\nabla_{\mu}\theta^{(k)}$.
However horizons coincide on~$\cal{S}$ and the wormhole would initially be traversable if both normals are spacelike
\begin{eqnarray}
  l^{\mu}\nabla_{\mu}\theta^{(l)}&>&0\label{lflareout}\\
  k^{\mu}\nabla_{\mu}\theta^{(k)}&>&0\label{kflareout}
\end{eqnarray}
i.e. trapping horizons are timelike hypersurfaces in the vicinity of~$\cal{S}$. 
The inequalities~\eqref{lflareout} and~\eqref{kflareout} are known as the flare-out conditions.

On the other hand, using the Raychaudhuri equation together with~\eqref{letensor} and~\eqref{ketensor} we can
write
\begin{widetext}
  \begin{eqnarray}
    l^{\mu}\nabla_{\mu}\theta^{(l)} &=& -\theta^{(l)}_{\mu\nu}\theta_{(l)}^{\mu\nu}-8\pi l^2_{\mathrm{P}}T_{\mu\nu}l^{\mu}l^{\nu}=-\frac{1}{2}H_{\mu\nu}H^{\mu\nu}-8\pi l^2_{\mathrm{P}}T_{\mu\nu}l^{\mu}l^{\nu},\\
     k^{\mu}\nabla_{\mu}\theta^{(k)} &=& -\theta^{(k)}_{\mu\nu}\theta_{(k)}^{\mu\nu}-8\pi l^2_{\mathrm{P}}T_{\mu\nu}k^{\mu}k^{\nu}= -\frac{1}{2}H_{\mu\nu}H^{\mu\nu}-8\pi l^2_{\mathrm{P}}T_{\mu\nu}k^{\mu}k^{\nu}.
  \end{eqnarray}
\end{widetext}
Here~\mbox{$H_{\mu\nu}=-q^{\alpha}_{\mu}q^{\beta}_{\nu}D_{\alpha}s_{\beta}$} is the extrinsic curvature of~$\cal{S}$ in~$\Sigma$.
However, since the throat is a minimal surface it is a totally geodesic manifold
and therefore~\mbox{$H_{\mu\nu}=0$}. Eventually, if the wormhole supported by phantom scalar
field we obtain on the throat
\begin{subequations}
\begin{eqnarray}
  l^{\mu}\nabla_{\mu}\theta^{(l)} &=& 4\pi l^2_{\mathrm{P}}(s^{\mu}D_{\nu}\phi)^2>0,\\
  k^{\mu}\nabla_{\mu}\theta^{(k)} &=& 4\pi l^2_{\mathrm{P}}(s^{\mu}D_{\nu}\phi)^2>0,
\end{eqnarray}
\label{flareout}
\end{subequations}
This the main result of the present section. It has just been shown that wormholes discussed in the paper
are initially traversable.

Note that the addition of the electric field does not change~\eqref{flareout}. Indeed, for the electric field we have
\begin{equation}
  T_{\mu\nu}l^{\mu}l^{\nu}=T_{\mu\nu}k^{\mu}k^{\nu}=\frac{1}{8\pi}\left[-(E_{\mu}s^{\mu})^2+E_{\mu}E^{\mu}\right]
\end{equation}
On the throat, however, the electric field is directed along the normal. Therefore, the expression in the brackets vanishes.

\bibliography{refs}

\providecommand{\noopsort}[1]{}\providecommand{\singleletter}[1]{#1}%
\begin{thebibliography}{24}%
\makeatletter
\providecommand \@ifxundefined [1]{%
 \@ifx{#1\undefined}
}%
\providecommand \@ifnum [1]{%
 \ifnum #1\expandafter \@firstoftwo
 \else \expandafter \@secondoftwo
 \fi
}%
\providecommand \@ifx [1]{%
 \ifx #1\expandafter \@firstoftwo
 \else \expandafter \@secondoftwo
 \fi
}%
\providecommand \natexlab [1]{#1}%
\providecommand \enquote  [1]{``#1''}%
\providecommand \bibnamefont  [1]{#1}%
\providecommand \bibfnamefont [1]{#1}%
\providecommand \citenamefont [1]{#1}%
\providecommand \href@noop [0]{\@secondoftwo}%
\providecommand \href [0]{\begingroup \@sanitize@url \@href}%
\providecommand \@href[1]{\@@startlink{#1}\@@href}%
\providecommand \@@href[1]{\endgroup#1\@@endlink}%
\providecommand \@sanitize@url [0]{\catcode `\\12\catcode `\$12\catcode
  `\&12\catcode `\#12\catcode `\^12\catcode `\_12\catcode `\%12\relax}%
\providecommand \@@startlink[1]{}%
\providecommand \@@endlink[0]{}%
\providecommand \url  [0]{\begingroup\@sanitize@url \@url }%
\providecommand \@url [1]{\endgroup\@href {#1}{\urlprefix }}%
\providecommand \urlprefix  [0]{URL }%
\providecommand \Eprint [0]{\href }%
\providecommand \doibase [0]{http://dx.doi.org/}%
\providecommand \selectlanguage [0]{\@gobble}%
\providecommand \bibinfo  [0]{\@secondoftwo}%
\providecommand \bibfield  [0]{\@secondoftwo}%
\providecommand \translation [1]{[#1]}%
\providecommand \BibitemOpen [0]{}%
\providecommand \bibitemStop [0]{}%
\providecommand \bibitemNoStop [0]{.\EOS\space}%
\providecommand \EOS [0]{\spacefactor3000\relax}%
\providecommand \BibitemShut  [1]{\csname bibitem#1\endcsname}%
\let\auto@bib@innerbib\@empty
\bibitem [{\citenamefont {Visser}(1995)}]{visser1}%
  \BibitemOpen
  \bibfield  {author} {\bibinfo {author} {\bibfnamefont {M.}~\bibnamefont
  {Visser}},\ }\href@noop {} {\emph {\bibinfo {title} {Lorentzian wormholes:
  From Einstein to Hawking}}}\ (\bibinfo  {publisher} {AIP},\ \bibinfo {year}
  {1995})\BibitemShut {NoStop}%
\bibitem [{\citenamefont {M.~S.~Morris}(1988)}]{thorne1}%
  \BibitemOpen
  \bibfield  {author} {\bibinfo {author} {\bibfnamefont {K.~S.~Thorne}\
  \bibnamefont {M.~S.~Morris}},\ }\href@noop {} {\bibfield  {journal} {\bibinfo
   {journal} {Am. J. Phys.}\ }\textbf {\bibinfo {volume} {56(5)}},\ \bibinfo
  {pages} {395} (\bibinfo {year} {1988})}\BibitemShut {NoStop}%
\bibitem [{\citenamefont {{Friedman}}\ \emph {et~al.}(1990)\citenamefont
  {{Friedman}}, \citenamefont {{Morris}}, \citenamefont {{Novikov}},
  \citenamefont {{Echeverria}}, \citenamefont {{Klinkhammer}}, \citenamefont
  {{Thorne}},\ and\ \citenamefont {{Yurtsever}}}]{thorne2}%
  \BibitemOpen
  \bibfield  {author} {\bibinfo {author} {\bibfnamefont {J.}~\bibnamefont
  {{Friedman}}}, \bibinfo {author} {\bibfnamefont {M.~S.}\ \bibnamefont
  {{Morris}}}, \bibinfo {author} {\bibfnamefont {I.~D.}\ \bibnamefont
  {{Novikov}}}, \bibinfo {author} {\bibfnamefont {F.}~\bibnamefont
  {{Echeverria}}}, \bibinfo {author} {\bibfnamefont {G.}~\bibnamefont
  {{Klinkhammer}}}, \bibinfo {author} {\bibfnamefont {K.~S.}\ \bibnamefont
  {{Thorne}}}, \ and\ \bibinfo {author} {\bibfnamefont {U.}~\bibnamefont
  {{Yurtsever}}},\ }\href@noop {} {\bibfield  {journal} {\bibinfo  {journal}
  {\prd}\ }\textbf {\bibinfo {volume} {42}},\ \bibinfo {pages} {1915} (\bibinfo
  {year} {1990})}\BibitemShut {NoStop}%
\bibitem [{\citenamefont {N.~O~Murchadha}(1974{\natexlab{a}})}]{york1}%
  \BibitemOpen
  \bibfield  {author} {\bibinfo {author} {\bibfnamefont {J.W.~York}\
  \bibnamefont {N.~O~Murchadha}},\ }\href@noop {} {\bibfield  {journal}
  {\bibinfo  {journal} {\prd}\ }\textbf {\bibinfo {volume} {10}},\ \bibinfo
  {pages} {428} (\bibinfo {year} {1974}{\natexlab{a}})}\BibitemShut {NoStop}%
\bibitem [{\citenamefont {N.~O~Murchadha}(1974{\natexlab{b}})}]{york2}%
  \BibitemOpen
  \bibfield  {author} {\bibinfo {author} {\bibfnamefont {J.W.~York}\
  \bibnamefont {N.~O~Murchadha}},\ }\href@noop {} {\bibfield  {journal}
  {\bibinfo  {journal} {\prd}\ }\textbf {\bibinfo {volume} {10}},\ \bibinfo
  {pages} {437} (\bibinfo {year} {1974}{\natexlab{b}})}\BibitemShut {NoStop}%
\bibitem [{\citenamefont {J.~A.~Isenberg}\ and\ \citenamefont
  {York}(1976)}]{york3}%
  \BibitemOpen
  \bibfield  {author} {\bibinfo {author} {\bibfnamefont {N.~O~Murchadha}\
  \bibnamefont {J.~A.~Isenberg}}\ and\ \bibinfo {author} {\bibfnamefont {J.W.}\
  \bibnamefont {York}},\ }\href@noop {} {\bibfield  {journal} {\bibinfo
  {journal} {\prd}\ }\textbf {\bibinfo {volume} {13}},\ \bibinfo {pages} {1532}
  (\bibinfo {year} {1976})}\BibitemShut {NoStop}%
\bibitem [{\citenamefont {Lindquist}(1963)}]{lindquist1}%
  \BibitemOpen
  \bibfield  {author} {\bibinfo {author} {\bibfnamefont {R.~W.}\ \bibnamefont
  {Lindquist}},\ }\href@noop {} {\bibfield  {journal} {\bibinfo  {journal} {J.
  Math. Phys.}\ }\textbf {\bibinfo {volume} {4}},\ \bibinfo {pages} {938}
  (\bibinfo {year} {1963})}\BibitemShut {NoStop}%
\bibitem [{\citenamefont {D.R.~Brill}(1963)}]{lindquist2}%
  \BibitemOpen
  \bibfield  {author} {\bibinfo {author} {\bibfnamefont {R.~W.~Lindquist}\
  \bibnamefont {D.R.~Brill}},\ }\href@noop {} {\bibfield  {journal} {\bibinfo
  {journal} {Phys.\ Rev.}\ }\textbf {\bibinfo {volume} {131}},\ \bibinfo
  {pages} {471} (\bibinfo {year} {1963})}\BibitemShut {NoStop}%
\bibitem [{\citenamefont {Misner}(1963)}]{misner2}%
  \BibitemOpen
  \bibfield  {author} {\bibinfo {author} {\bibfnamefont {C.~W.}\ \bibnamefont
  {Misner}},\ }\href@noop {} {\bibfield  {journal} {\bibinfo  {journal} {Ann.
  of Phys..}\ }\textbf {\bibinfo {volume} {24}},\ \bibinfo {pages} {102}
  (\bibinfo {year} {1963})}\BibitemShut {NoStop}%
\bibitem [{\citenamefont {Misner}(1960)}]{misner1}%
  \BibitemOpen
  \bibfield  {author} {\bibinfo {author} {\bibfnamefont {C.~W.}\ \bibnamefont
  {Misner}},\ }\href@noop {} {\bibfield  {journal} {\bibinfo  {journal} {Phys.\
  Rev.}\ }\textbf {\bibinfo {volume} {118}},\ \bibinfo {pages} {1110} (\bibinfo
  {year} {1960})}\BibitemShut {NoStop}%
\bibitem [{\citenamefont {Ellis}(1973)}]{ellis1}%
  \BibitemOpen
  \bibfield  {author} {\bibinfo {author} {\bibfnamefont {H.~G.}\ \bibnamefont
  {Ellis}},\ }\href@noop {} {\bibfield  {journal} {\bibinfo  {journal} {J.
  Math. Phys.}\ }\textbf {\bibinfo {volume} {14}},\ \bibinfo {pages} {395}
  (\bibinfo {year} {1973})}\BibitemShut {NoStop}%
\bibitem [{\citenamefont {Bronnikov}(1973)}]{bronnikov1}%
  \BibitemOpen
  \bibfield  {author} {\bibinfo {author} {\bibfnamefont {K.~A.}\ \bibnamefont
  {Bronnikov}},\ }\href@noop {} {\bibfield  {journal} {\bibinfo  {journal}
  {Acta Phys. Pol. B}\ }\textbf {\bibinfo {volume} {4}},\ \bibinfo {pages}
  {251} (\bibinfo {year} {1973})}\BibitemShut {NoStop}%
\bibitem [{\citenamefont {Giulini}(1990)}]{giulini1}%
  \BibitemOpen
  \bibfield  {author} {\bibinfo {author} {\bibfnamefont {D.}~\bibnamefont
  {Giulini}},\ }\href@noop {} {\bibfield  {journal} {\bibinfo  {journal}
  {Class. Quantum Grav.}\ }\textbf {\bibinfo {volume} {7}},\ \bibinfo {pages}
  {1271} (\bibinfo {year} {1990})}\BibitemShut {NoStop}%
\bibitem [{\citenamefont {Alqubierre}(2008)}]{alqubierre1}%
  \BibitemOpen
  \bibfield  {author} {\bibinfo {author} {\bibfnamefont {M.}~\bibnamefont
  {Alqubierre}},\ }\href@noop {} {\emph {\bibinfo {title} {Introduction to 3+1
  Numerical Relativity}}}\ (\bibinfo  {publisher} {Oxford University Press},\
  \bibinfo {year} {2008})\BibitemShut {NoStop}%
\bibitem [{\citenamefont {S.~A.~Hayward}(2002)}]{hayward2}%
  \BibitemOpen
  \bibfield  {author} {\bibinfo {author} {\bibfnamefont {H.~Shinkai}\
  \bibnamefont {S.~A.~Hayward}},\ }\href@noop {} {\bibfield  {journal}
  {\bibinfo  {journal} {\prd}\ }\textbf {\bibinfo {volume} {66}},\ \bibinfo
  {pages} {044005} (\bibinfo {year} {2002})}\BibitemShut {NoStop}%
\bibitem [{\citenamefont {Gonzalez}\ \emph
  {et~al.}(2009{\natexlab{a}})\citenamefont {Gonzalez}, \citenamefont
  {Guzman},\ and\ \citenamefont {Sarbach}}]{gonzalez2}%
  \BibitemOpen
  \bibfield  {author} {\bibinfo {author} {\bibfnamefont {J.~A.}\ \bibnamefont
  {Gonzalez}}, \bibinfo {author} {\bibfnamefont {F.~S.}\ \bibnamefont
  {Guzman}}, \ and\ \bibinfo {author} {\bibfnamefont {O.}~\bibnamefont
  {Sarbach}},\ }\href@noop {} {\bibfield  {journal} {\bibinfo  {journal}
  {Class. Quant. Grav.}\ }\textbf {\bibinfo {volume} {26}},\ \bibinfo {pages}
  {015010} (\bibinfo {year} {2009}{\natexlab{a}})}\BibitemShut {NoStop}%
\bibitem [{\citenamefont {Gonzalez}\ \emph
  {et~al.}(2009{\natexlab{b}})\citenamefont {Gonzalez}, \citenamefont
  {Guzman},\ and\ \citenamefont {Sarbach}}]{gonzalez3}%
  \BibitemOpen
  \bibfield  {author} {\bibinfo {author} {\bibfnamefont {J.~A.}\ \bibnamefont
  {Gonzalez}}, \bibinfo {author} {\bibfnamefont {F.~S.}\ \bibnamefont
  {Guzman}}, \ and\ \bibinfo {author} {\bibfnamefont {O.}~\bibnamefont
  {Sarbach}},\ }\href@noop {} {\bibfield  {journal} {\bibinfo  {journal}
  {Class. Quant. Grav.}\ }\textbf {\bibinfo {volume} {26}},\ \bibinfo {pages}
  {015011} (\bibinfo {year} {2009}{\natexlab{b}})}\BibitemShut {NoStop}%
\bibitem [{\citenamefont {Hayward}(1999)}]{hayward1}%
  \BibitemOpen
  \bibfield  {author} {\bibinfo {author} {\bibfnamefont {S.~A.}\ \bibnamefont
  {Hayward}},\ }\href@noop {} {\bibfield  {journal} {\bibinfo  {journal} {Int.
  J. Mod. Phys. D}\ }\textbf {\bibinfo {volume} {8}},\ \bibinfo {pages} {373}
  (\bibinfo {year} {1999})}\BibitemShut {NoStop}%
\bibitem [{\citenamefont {Matos}(2010)}]{matos1}%
  \BibitemOpen
  \bibfield  {author} {\bibinfo {author} {\bibfnamefont {T.}~\bibnamefont
  {Matos}},\ }\href@noop {} {\bibfield  {journal} {\bibinfo  {journal} {Gen.
  Rel. Grav.}\ }\textbf {\bibinfo {volume} {42}},\ \bibinfo {pages}
  {1969--1990} (\bibinfo {year} {2010})}\BibitemShut {NoStop}%
\bibitem [{\citenamefont {Gonzalez}\ \emph
  {et~al.}(2009{\natexlab{c}})\citenamefont {Gonzalez}, \citenamefont
  {Guzman},\ and\ \citenamefont {Sarbach}}]{gonzalez1}%
  \BibitemOpen
  \bibfield  {author} {\bibinfo {author} {\bibfnamefont {J.~A.}\ \bibnamefont
  {Gonzalez}}, \bibinfo {author} {\bibfnamefont {F.~S.}\ \bibnamefont
  {Guzman}}, \ and\ \bibinfo {author} {\bibfnamefont {O.}~\bibnamefont
  {Sarbach}},\ }\href@noop {} {\bibfield  {journal} {\bibinfo  {journal} {Phys.
  Rev. D}\ }\textbf {\bibinfo {volume} {80}},\ \bibinfo {pages} {024023}
  (\bibinfo {year} {2009}{\natexlab{c}})}\BibitemShut {NoStop}%
\bibitem [{\citenamefont {Ortin}(1995)}]{ortin1}%
  \BibitemOpen
  \bibfield  {author} {\bibinfo {author} {\bibfnamefont {T.}~\bibnamefont
  {Ortin}},\ }\href@noop {} {\bibfield  {journal} {\bibinfo  {journal} {Phys.\
  Rev. D}\ }\textbf {\bibinfo {volume} {52}},\ \bibinfo {pages} {3392}
  (\bibinfo {year} {1995})}\BibitemShut {NoStop}%
\bibitem [{\citenamefont {Tod}(1983)}]{tod1}%
  \BibitemOpen
  \bibfield  {author} {\bibinfo {author} {\bibfnamefont {K.~P.}\ \bibnamefont
  {Tod}},\ }\href@noop {} {\bibfield  {journal} {\bibinfo  {journal} {Proc.
  Roy. Soc. Lond. A}\ }\textbf {\bibinfo {volume} {388}},\ \bibinfo {pages}
  {457--477} (\bibinfo {year} {1983})}\BibitemShut {NoStop}%
\bibitem [{\citenamefont {K.~Nakao}(1993)}]{maeda1}%
  \BibitemOpen
  \bibfield  {author} {\bibinfo {author} {\bibfnamefont {K.~Maeda}\
  \bibnamefont {K.~Nakao}, \bibfnamefont {K.~Yamamoto}},\ }\href@noop {}
  {\bibfield  {journal} {\bibinfo  {journal} {\prd}\ }\textbf {\bibinfo
  {volume} {47}},\ \bibinfo {pages} {3203} (\bibinfo {year}
  {1993})}\BibitemShut {NoStop}%
\bibitem [{\citenamefont {J.~Yang}(2021)}]{yang1}%
  \BibitemOpen
  \bibfield  {author} {\bibinfo {author} {\bibfnamefont {H.~Huang}\
  \bibnamefont {J.~Yang}},\ }\href@noop {} {\bibfield  {journal} {\bibinfo
  {journal} {\prd}\ }\textbf {\bibinfo {volume} {104}},\ \bibinfo {pages}
  {084005} (\bibinfo {year} {2021})}\BibitemShut {NoStop}%
\end{thebibliography}%

\end{document}